# SIMPLE SOLUTION FOR DESIGNING THE PIECEWISE LINEAR SCALAR COMPANDING QUANTIZER FOR GAUSSIAN SOURCE

Jelena Nikolić[1], Zoran Perić[1], Lazar Velimirović[2]

[1]*Faculty of Electronic Engineering, University of Niš, Aleksandra Medvedeva 14, 18000 Niš, Serbia*
[2]*Mathematical Institute of the Serbian Academy of Sciences and Arts, Kneza Mihaila 36, 11001 Belgrade, Serbia*
*email: jelena.nikolic@elfak.ni.ac.rs, zoran.peric@elfak.ni.ac.rs,velimirovic.lazar@gmail.com*

**Abstract.** To overcome the difficulties in determining an inverse compressor function for a Gaussian source, which appear in designing the nonlinear optimal companding quantizers and also in the nonlinear optimal companding quantization procedure, in this paper a piecewise linear compressor function based on the first derivate approximation of the optimal compressor function is proposed. We show that the approximations used in determining the piecewise linear compressor function contribute to the simple solution for designing the novel piecewise linear scalar companding quantizer (PLSCQ) for a Gaussian source of unit variance. For the given number of segments, we perform optimization procedure in order to obtain optimal value of the support region threshold which maximizes the signal to quantization noise ratio (SQNR) of the proposed PLSCQ. We study how the SQNR of the considered PLSCQ depends on the number of segments and we show that for the given number of quantization levels, SQNR of the PLSCQ approaches the one of the nonlinear optimal companding quantizer with the increase of the number of segments. The presented features of the proposed PLSCQ indicate that the obtained model should be of high practical significance for quantization of signals having Gaussian probability density function.

## 1. Introduction

A great interest in quantization is generally motivated by the evolution to digital communications, i.e. more specifically, by the tradeoff between lowering the bit rate and maintaining the quality of the quantized signal. Many other constrains can be considered, such as complexity and delay [8]. In this paper there is a restriction with respect to the complexity since our goal is to provide a very simple quantizer solution having a smaller complexity than the widely used nonlinear companding quantizer.

For a fixed number of quantization levels $N$, or equivalently a fixed bit rate $R$ [bit/sample] = $\log_2 N$ [8], reproduction levels and partition regions or cells of a quantizer can be determined according to a different criteria. Generally, the primary goal of a quantizer design is to obtain minimal possible distortion, i.e. maximal possible signal to quantization noise ratio (SQNR). Lloyd and Max developed an algorithm for designing an optimal quantizer having a minimal possible distortion [4]. However, this algorithm is too time consuming for the large number of quantization levels we are interested in. One solution which overcomes this problem providing the performance close to the optimal one is defined by the nonlinear optimal companding quantizer model [8]. However, it is well known that designing nonlinear optimal companding quantizers for a Gaussian source is very complex due to the difficulties in determining the inverse optimal compressor function [7], [8], [11]-[13]. Also, from the aspect of hardware, it is very difficult to pair the characteristics of diodes that are used for the implementation of a compressor and an expandor [8]. Moreover, the software implementation of companding quantizers meets many difficulties. With these models, in accordance with the condition of the nearest neighbors, for the quantization of each sample, $N$ distortion estimate is carried out, so that for each input sample a complete search of the code book is performed [8]. Accordingly, there is an evident need for simplifying the design procedure of companding quantizers, where the goal is to preserve performance as much as possible. One of the manners to achieve this goal is based on the linearization of the compressor function and the resulting quantizers are known as piecewise linear scalar companding quantizers (PLSCQs).

In PLSCQ the support region consists of several segments, each of which containing several uniform quantization cells and uniformly distributed reproduction levels [8]. The fact that these quantizers are piecewise linear and hence, conceptually and implementationally simpler then the nonuniform quantizers [8] justifies their widespread application. For instance, to achieve high-quality quantized speech signals, the contemporary public switched telephone networks utilize the PLSCQ proposed by the G.711 Recommendation [3]. G.711 quantizer has the advantages of low complexity and delay with high-quality reproduced speech, but require a relatively high bit rate. Namely, G.711 Recommendation defines a symmetric PLSCQ by 8 bits of resolution ($R = 8$ bit/sample) and $L = 8$ positive segments increased in length by a factor of 2 for each successive segments having 16 cells [3], [8]. The G.711 quantizers, based on the piecewise linear approximation to the $A$-law and $\mu$-law compressor functions, divide the support region into a $2L = 16$ unequal segments with equal number of cells. Along with the support region partition, according to the mentioned piecewise linear compressor functions, there are some novel propositions of the support region partition, i.e. of the PLSCQ design. For instance, the robustness conditions of the PLSCQ based on a piecewise linear approximation to the optimal compressor function are analyzed in [2]. A comprehensive analysis of SQNR behavior in the wide range of variances for the PLSCQ designed for the Laplacian source of unit variance according to the piecewise linear approximation to the optimal compressor function is reported in [5]. Unlike the PLSCQ proposed in [5], where the number of cells is assumed to be constant per segments and where the segments are determined by the equidistant partition of the optimal compressor function, the number of cells per segments has been optimized in [7], for the case of a Gaussian source of unit variance. This contributes to the SQNR increase.

As reported in [7], one of the reasons of often considering the Gaussian source is that the first approximation to the short-time-averaged probability density function (PDF) of speech amplitudes is provided by the Gaussian PDF. Also, one reason for studying the Gaussian source is that it naturally arises in numerous applications. For example, the prediction error signal in a differential pulse-code modulation (DPCM) coder for moving pictures is well-modeled as Gaussian [10]. Discrete Fourier transform coefficients and holographic data can often considered to be the output of a Gaussian source [9]. Finally, since a properly chosen filtering technique applied to non-Gaussian source produces sequences which are approximately independent and Gaussian, a quantizer designed for the Gaussian source can also be applied to other sources, providing the similar performance [6].

The difference between the quantizer model we propose in this paper and the one described in [7] is in the manner of determining the number of cells per segments and in the realization structure. With the quantizer model proposed in this paper the number of cells per segments is determined according to the piecewise linear compressor function, whereas with the quantizer model described in [7], the method of Lagrange multipliers is used in order to optimize the number of cells per segments. In fact, the quantizer described in [7] is not a piecewise linear scalar companding quantizer, as the one we propose, but instead a piecewise uniform scalar quantizer, which can be considered as a set of uniform quantizers, where the number of uniform quantizers is equal to the number of segments. What we propose in this paper is a novel model of PLSCQ having the piecewise linear compressor function determined by the first derivate approximation of the optimal compressor function at the point on the middle of the segments. The novel model is very simple to design, even for the case of a Gaussian source, since it does not require determining the solutions of the complex system of integral equations, as in the case of nonlinear optimal companding quantizers. In addition, due to the piecewise linear property of the compressor function, with this model there are no difficulties with pairing the characteristics of diodes that are used for the implementation of a compressor and an expandor.

The rest of this paper is organized as follows. Section 2 presents a detailed description of the novel PLSCQ. The achieved numerical results for the Gaussian source of unit variance are discussed in Section 3. Finally, Section 4 is devoted to the conclusions which summarise the contribution achieved in the paper.

## 2. Design of the novel piecewise linear scalar companding quantizer for Gaussian source

The optimal compressor function $c(x)$ by which the maximum of SQNR is achieved for the reference variance $\sigma^2$ of an input signal $x$ is defined as [8]:

$$c(x) = \begin{cases} x_{\max} \dfrac{\int_0^x p^{1/3}(x)dx}{\int_0^{x_{\max}} p^{1/3}(x)dx}, & 0 \leq x \leq x_{\max} \\ -x_{\max} \dfrac{\int_x^0 p^{1/3}(x)dx}{\int_{-x_{\max}}^0 p^{1/3}(x)dx}, & -x_{\max} \leq x \leq 0 \end{cases}, \quad (1)$$

where $x_{\max}$ denotes the support region threshold of the optimal companding quantizer. As already mentioned, there are some difficulties when designing the companding quantizers in the case of a Gaussian source. Accordingly, our goal is to obtain a piecewise linear compressor function which provides the simple quantizer design even for the case of a Gaussian source.

In the rest of the paper we assume symmetry about zero in the PLSCQ design. This symmetry is an intuitively expected result when the input has a PDF that is symmetrical about zero. The Gaussian PDF, we consider here, is indeed symmetrical about zero. Namely, without loss of generality, we assume that information source is Gaussian source with memoryless property, the unit variance and zero mean value. The PDF of this source is given by [8]:

$$p(x) = \frac{1}{\sqrt{2\pi}} \exp\left(-\frac{x^2}{2}\right). \quad (2)$$

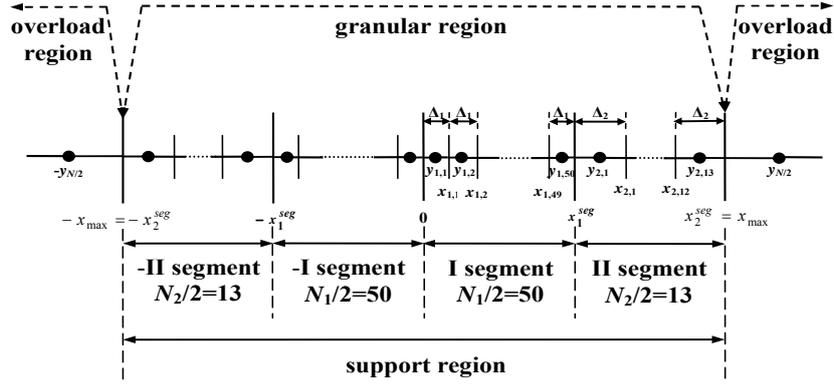

**Fig. 1.** Support region partition of the proposed PLSCQ for the case where the total number of levels is $N = 128$ and $L = 2$: $x_L^{seg} = x_{max}$ is support region threshold; $x_i^{seg}$, $i = 0, 1,..., L$, are segment thresholds; $\Delta_i$, $i = 1,..., L$, are cell lengths; $x_{i,j}$, $i = 1,..., L$, $j = 1,..., N_i$, are cell thresholds; $y_{i,j}$, $i = 1,..., L$, $j = 1,..., N_i$, are reproduction levels.

We propose a PLSCQ with $N$ levels and $2L$ segments, having equidistant segment thresholds determined by:

$$x_i^{seg} = i\frac{x_{max}}{L}, \quad x_{-i}^{seg} = -x_i^{seg}, \quad i = 0,1,...,L, \quad (3)$$

where obviously for the support region threshold it holds $x_{max} = x_L^{seg}$. Let $N_i/2$ be the number of cells within the corresponding $i$-th segment (see Fig. 1). Cell lengths of the considered PLSCQ are equal within the segment whereas they may be different from segment to segment:

$$\Delta_i = \frac{x_i^{seg} - x_{i-1}^{seg}}{N_i/2} = \frac{2 x_L^{seg}}{L N_i}, \quad i = 1,..., L. \quad (4)$$

As mentioned above, we assume symmetry in the PLSCQ design and accordingly, we define the parameters of the PLSCQ that correspond to the $L$ positive segments. Denote by $y_{i,j}$ the $j$-th reproduction level within the $i$-th segment $(x_{i-1}^{seg}, x_i^{seg}]$. In the case where the current amplitude value of the input signal falls in the $j$-th cell within the $i$-th segment $(x_{i,j-1}, x_{i,j}]$, where:

$$x_{i,j} = x_{i-1}^{seg} + j \Delta_i, \quad i = 1,..., L, \quad j = 1,..., N_i, \quad (5)$$

the quantization rule provides its coping onto the near allowed value $y_{i,j}$, defined by:

$$y_{i,j} = x_{i-1}^{seg} + \frac{(2j-1)}{2}\Delta_i, \quad i = 1,..., L, \quad j = 1,..., N_i. \quad (6)$$

In other words, the cell thresholds and the reproduction levels of the considered PLSCQ are defined as $x_k = x_{i,j}$, $y_k = y_{i,j}$, $i = 1,..., L$, $j = 1,..., N_i$, $k = 1,..., (N - 2)/2$, where for the outermost reproduction level, as in [11], the centroid condition is assumed in the design process:

$$y_{N/2} = \frac{\int_{x_L^{seg}}^{\infty} x\, p(x) dx}{\int_{x_L^{seg}}^{\infty} p(x) dx}. \quad (7)$$

Our model is based on the first derivate approximation of the optimal compressor function (1) at the points on the middle of the segments. These points are given by:

$$s_i = \frac{(2i-1) x_L^{seg}}{2L}, \quad i = 1,..., L. \quad (8)$$

The first derivate of the optimal compressor function (1) at $s_i$, $i = 1,..., L$, for the case $0 \le x \le x_L^{seg}$, is:

$$c'(s_i) = \frac{x_L^{seg} p^{1/3}(s_i)}{\int_0^{x_L^{seg}} p^{1/3}(x) dx}, \quad i = 1,..., L, \quad (9)$$

where $x_{max}$ is substituted by $x_L^{seg}$. Obviously, we have $L$ different slopes of the piecewise linear compressor function, which are determined by the $c'(s_i)$, $i = 1,..., L$. In order to define a novel piecewise linear compressor function we can assume the following approximations:

$$\int_0^{x_{i-1}^{seg}} p^{1/3}(x) dx \approx \sum_{j=1}^{i-1} p^{1/3}(s_j)\frac{x_L^{seg}}{L}, \quad i = 2,..., L, \quad (10)$$

$$\int_{x_{i-1}^{seg}}^{x} p^{1/3}(x) dx \approx (x - x_{i-1}^{seg}) p^{1/3}(s_i), \quad x_{i-1}^{seg} < x \le x_i^{seg}, \quad (11)$$

$$\int_0^{x_L^{seg}} p^{1/3}(x) dx \approx \sum_{i=1}^{L} p^{1/3}(s_i)\frac{x_L^{seg}}{L}, \quad (12)$$

that yields:

$$\int_0^{x} p^{1/3}(x) dx \approx \sum_{j=1}^{i-1} p^{1/3}(s_j)\frac{x_L^{seg}}{L} + (x - x_{i-1}^{seg}) p^{1/3}(s_i), \quad (13)$$

$$c_1(x) = x_L^{seg} \frac{(x - x_0^{seg}) p^{1/3}(s_1)}{\sum_{i=1}^{L} p^{1/3}(s_i)\frac{x_L^{seg}}{L}}, \quad (14)$$

$$c_i(x) = x_L^{seg} \frac{\sum_{j=1}^{i-1} p^{1/3}(s_j)\frac{x_L^{seg}}{L} + (x - x_{i-1}^{seg}) p^{1/3}(s_i)}{\sum_{i=1}^{L} p^{1/3}(s_i)\frac{x_L^{seg}}{L}}, \quad i = 2,..., L. \quad (15)$$

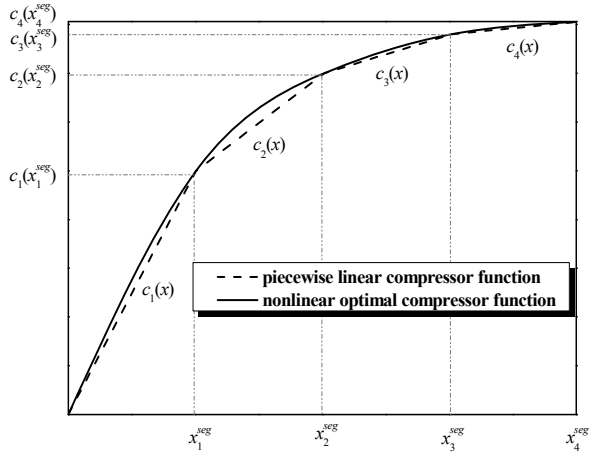

**Fig. 2.** Piecewise linear compressor function and nonlinear optimal compressor function for the number of segments $2L = 8$.

$c_i(x)$, $i = 1,\ldots,L$, defined by (14) and (15) is a novel piecewise linear compressor function (see Fig. 2), which is a continuous function because it holds:

$$c_i\left(x_i^{seg}\right) = c_{i+1}\left(x_i^{seg}\right), \quad i = 1,\ldots,L-1. \qquad (16)$$

We assume that the total number of cells per segments in the first quadrant is:

$$\sum_{i=1}^{L} \frac{N_i}{2} = \frac{N-2}{2}. \qquad (17)$$

Accordingly, we can define the following:

$$\Delta = \frac{c_L\left(x_L^{seg}\right)}{(N-2)/2}, \qquad (18)$$

$$c_i\left(x_i^{seg}\right) - c_i\left(x_{i-1}^{seg}\right) = \frac{N_i}{2}\Delta, \quad i = 1,\ldots,L. \qquad (19)$$

Combining (18) and (19) enables determining the number of cells per segments for the proposed PLSCQ model:

$$\frac{N_i}{2} = \frac{(N-2)}{2} \frac{\left(c_i\left(x_i^{seg}\right) - c_i\left(x_{i-1}^{seg}\right)\right)}{c_L\left(x_L^{seg}\right)}. \qquad (20)$$

The granular distortion $D_g$ and the overload distortion $D_o$ of the proposed PLSCQ can be determined by using the basic definition for the granular distortion of the PLSCQ given in [8]:

$$D_g = 2\sum_{i=1}^{L} \frac{\Delta_i^2}{12} P_i, \qquad (21)$$

$$P_i = \int_{x_{i-1}^{seg}}^{x_i^{seg}} p(x)dx = \frac{1}{2}\left[erf\left(\frac{x_i^{seg}}{\sqrt{2}}\right) - erf\left(\frac{x_{i-1}^{seg}}{\sqrt{2}}\right)\right], \qquad (22)$$

where $\Delta_i$ is given by (4), and by using the following closed-form formula derived in [11]:

$$D_o = 2\int_{x_L^{seg}}^{\infty} (x - y_{N/2})^2 p(x)dx = \sqrt{\frac{2}{\pi}} \frac{1}{\left(x_L^{seg}\right)^3} \exp\left(-\frac{\left(x_L^{seg}\right)^2}{2}\right). \qquad (23)$$

By determining the total distortion $D$, that is equal to the sum of the granular distortion $D_g$ (21) and the overload distortion $D_o$ (23), one can also determine the SQNR of the proposed PLSCQ by using the basic definition for the SQNR [8]:

$$\text{SQNR [dB]} = 10\log_{10}\left(\frac{1}{D_g + D_o}\right) = 10\log_{10}\left(\frac{1}{D}\right). \qquad (24)$$

In this paper, the analysis of numerical results is conducted using SQNR rather than distortion.

### 3. Numerical results

Numerical results presented in this section are obtained for the case $L = 1$, $L = 2$, $L = 4$ and $L = 8$, and for number of quantization levels $N = 128$, as it has been observed in [5]. By designing the PLSCQ for the Gaussian source of unit variance and for the support region threshold defined as in [11]:

$$x_{\max} = x_L^{seg} = \sqrt{6\ln N}\left[1 - \frac{\ln(\ln(N))}{4\ln(N)} - \frac{\ln(3\sqrt{\pi})}{2\ln(N)}\right], \qquad (25)$$

we have determined the SQNR characteristic of the proposed PLSCQ, denoted as PLSCQ[(1)] in Fig. 3. In addition, for the same $N = 128$, by assuming different support region thresholds, we have numerically determined the values of the optimal support region thresholds $x_1^{seg} = 3.50$, $x_2^{seg} = 3.80$, $x_4^{seg} = 3.98$ and $x_8^{seg} = 4.03$ that minimize the PLSCQ distortion (i.e. that maximize SQNR) for the cases when $L = 1$, $L = 2$, $L = 4$ and $L = 8$, respectively. By assuming these support region thresholds we have determined the SQNR characteristic, denoted as PLSCQ[(2)] in Fig. 3. Namely, one way to determine how well the nonlinear optimal companding quantizer and the PLSCQ match is to compare the SQNR characteristics for the same input statistics and the number of quantization levels, but for a different number of segments (see Fig. 3). Regarding such obtained SQNR characteristics one can conclude that starting from $L = 1$ to $L = 8$ the SQNR characteristics of the PLSCQ approaches to the one of the nonlinear optimal companding quantizer, where for $L < 4$, the PLSCQ[(2)] outperforms the PLSCQ[(1)]. Observe that in the case of $L = 1$, the considered PLSCQ is a uniform quantizer, as the one reported in [1]. This notice justifies the large SQNR degradation of 3.17 dB and 2.54 dB, which we have observed for $L = 1$ in respect to the nonlinear optimal companding quantizer, in the case of the PLSCQ[(1)] and PLSCQ[(2)], respectively. In addition, observe that by increasing $L$ to $L = 2$, for the same $N = 128$, the SQNR of the PLSCQ[(1)] and PLSCQ[(2)] is increased for 1.93 dB and 1.51 dB, respectively. Eventually, for $L = 8$, the SQNR of the proposed PLSCQ approaches the one of the nonlinear optimal companding quantizer, where the difference in SQNR amounts to 0.12 dB. For the given $N = 128$ and $L = 8$, with the PLSCQ designed in [5] for the Laplacian source of unit variance, we have achieved much greater difference in SQNR that amounts to 1 dB. Moreover, for $N = 128$ and $L = 8$, the SQNR of the PLSCQ[(2)] agrees well with that of the quantizer proposed in [7]. However, the goal in [7] was to provide the gain in SQNR when compared to the uniform quantizer, where in this paper we have proposed a simple solution for designing the PLSCQ where we have managed to achieve the SQNR very close to the one of the nonlinear optimal companding quantizer.

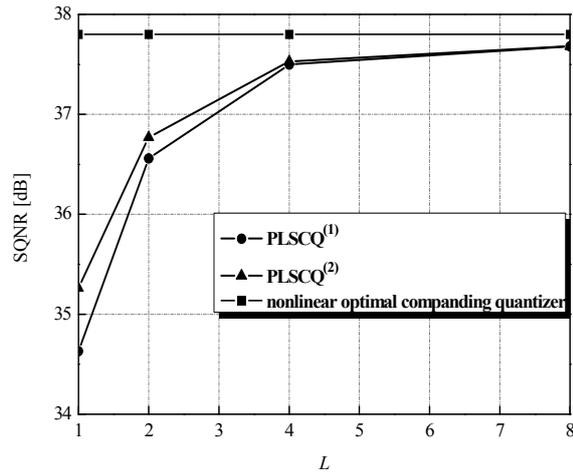

**Fig. 3.** Dependency of SQNR on the number of segments for the PLSCQ[(1)] and PLSCQ[(2)] and the nonlinear optimal companding quantizer

## 4. Summary and conclusion

In this paper, a novel solution of the piecewise linear scalar companding quantizer (PLSCQ) designed for the Gaussian source of unit variance has been proposed. The novel PLSCQ has been designed according the piecewise linear compressor function which is determined by the first derivate approximation of the optimal compressor function at the points on the middle of the segments. It has been shown that designing the proposed PLSCQ for a Gaussian source is very simple because determining the solutions of the complex system of integral equations is not required as it is in the case with designing the nonlinear optimal companding quantizers for the same source. Moreover, it has been shown that for the observed number of quantization levels, the SQNR of the proposed PLSCQ approaches very close to the one of the nonlinear optimal companding quantizer, where better convergency is achieved in the case when the support region threshold of the PLSCQ is optimized than in the case when it is formula-evaluated for the nonlinear optimal companding quantizer. All aforementioned points out the reasons why our PLSCQ model is suitable for use in many applications for the quantization of signals with Gaussian distribution.

## 5. Acknowledgments

This work is partially supported by Serbian Ministry of Education and Science through Mathematical Institute of Serbian Academy of Sciences and Arts (Project III44006) and by Serbian Ministry of Education and Science (Project TR32035).